\documentclass[aps,prl,twocolumn,superscriptaddress,showpacs]{revtex4}
\usepackage{hyperref}
\usepackage[utf8]{inputenc}
\usepackage[english]{babel}
\usepackage{graphicx}
\usepackage{amsmath}
\usepackage{amsfonts}
\usepackage{amssymb}
\usepackage{ulem}
 
\begin{document}

\title{Magnetoresistance in quasi-one dimensional Weyl semimetal (TaSe$_4$)$_2$I}

\author{I.A. Cohn}
\affiliation{Kotelnikov Institute of Radioengineering and Electronics of RAS, Mokhovaya 11, bld. 7, 125009 Moscow, Russia}

\author{S.G. Zybtsev}
\affiliation{Kotelnikov Institute of Radioengineering and Electronics of RAS, Mokhovaya 11, bld. 7, 125009 Moscow, Russia}
\author{A.P. Orlov}
\affiliation{Kotelnikov Institute of Radioengineering and Electronics of RAS, Mokhovaya 11, bld. 7, 125009 Moscow, Russia}
\author{S.V. Zaitsev-Zotov}
\email[]{serzz@cplire.ru} 
\affiliation{Kotelnikov Institute of Radioengineering and Electronics of RAS, Mokhovaya 11, bld. 7, 125009 Moscow, Russia}
\date{\today}

\begin{abstract}
Magnetic field effect on linear and nonlinear conductivity in a quasi-one-dimensional Weyl semimetal with a charge density wave (CDW) (TaSe$_4$)$_2$I is studied. Longitudinal magnetoresistance in all known regimes of CDW motion (linear conduction, creep, sliding, “Fr\"ohlich superconductivity”) is small, positive and do not exceed a fraction of per cent. Similar magnetotransport measurements were performed in samples profiled by focused ion beams is such a way that motion of the CDW in them is accompanied by phase slip of the CDW. In such samples, a peak-like non-parabolic negative magnetoresistance is observed in relatively small magnetic fields $B \lesssim 4$ T in the nonlinear conduction regime in both longitudinal and transverse geometries. Our results differ significantly from ones obtained earlier and raise the question concerning conditions for observing the axion anomaly in Weyl semimetals in the Peierls state.
\end{abstract}
\maketitle
Topological materials are currently some of the most interesting and intensively studied objects in solid state physics. Of great interest are Weyl semimetals, the transport properties of which are determined by the massless Weyl fermions. One of the most remarkable effects arising in such materials is the appearance of a chiral anomaly, which manifests itself as a negative longitudinal magnetoresistance (MR)  \cite{Nielsen,Son,Burkov}. At present, negative longitudinal MR has been observed in many Weyl and Dirac semimetals, such as  TaAs \cite{TaAs}, Cd$_3$As$_2$ \cite{CdAs}, {\it etc.} 

(TaSe$_4$)$_2$I has a monoclinic unit cell (space group I422) which consists of TaSe$_4$ chains with helical symmetry that are placed in the middle of the faces and separated by chains of iodine atoms  \cite{CDWreview}. The calculations of the band structure (TaSe $_4$)$_ 2$I show that under normal conditions this material is a semimetal due to the interaction between the TaSe$_4$ chains. This interaction leads to the splitting of the $d_{z^2}$ zones of tantalum into two intersecting near the boundary of the Brillouin zone at a distance of $0.44\pi /c$, where $c = 12.8$~\AA \ is the unit cell size along the chains \cite{Gressier}. Subsequent analysis shows that this material is a Weyl semimetal indeed in which numerous Weyl points are located at distances of 10-15 meV above and below the Fermi level and form pairs with the opposite chiral charge \cite{gooth,shi,li,zhang}.
 Moreover, (TaSe$_4$)$_2$I is a quasi-one-dimensional (quasi-1D) conductor with an incommensurate CDW forming at temperatures below the Peierls transition temperature $T_P \approx 248$ - 263 K \cite{wangmonceau,makigruner,forro,termoEDS,s1,s2,gooth,shi}. The Peierls gap covers the entire Fermi surface. This means the disappearance of the Weyl cones in the Peierls state. The conductivity anisotropy at room temperature is $ A\equiv \sigma_ \parallel / \sigma_\perp \approx 3 \times 10^2 $, where $\sigma_ \parallel$ and $\sigma_\perp$ conductivity along and across the chains, respectively and decreases to $ A\approx 10^2$ at $T <T_P$ \cite{termoEDS}.

In (TaSe$_4$)$_2$I, in the case when the magnetic field is applied parallel to the chains, and hence the line connecting the vertices of the Weyl cones, a negative longitudinal magnetoresistance (MR) is observed. Such an MR appears only in the nonlinear conductivity region in the electric field $E\approx 2$ - $3 E_T$, where $E_T$ is the threshold field for the onset of CDW sliding \cite{gooth}. The MR increases from 1-2\% at 120 K to almost an order at 80 K. Based on these observations, it was concluded in \cite{gooth} that in (TaSe$_4$)$_2$I, a chiral anomaly appears in the state with the Peierls gap.

The conductivity modes of quasi-one-dimensional conductors with the CDW can be classified as follows \cite{CDWreview,qq,FrSC}:

1) $E\ll E_T$. The CDW is pinned and does not contribute to nonlinear conductivity, since the potential barrier provided by pinning is too high to be overcome. Conductivity is linear and determined by quasiparticles (electrons and holes) thermally excited over the Peierls gap.

2) $E \lesssim E_T$. On approaching the threshold field, the pinning barrier gradually decreases and under certain conditions it becomes possible to observe the CDW creep that precedes sliding of the CDW \cite{qq}. In this mode, the nonlinear conductivity depends on the temperature in an activation manner. The activation energy decreases with approaching the threshold field and is not related to the conductivity of quasiparticles. The CDW creep leads to the disappearance of a sharp threshold for onset of sliding of the CDW.

3) $E > E_T$. The kinetics of the CDW is determined by the energy dissipation. Since the CDW cannot dissipate energy due to the energy gap, dissipation is provided by flows of quasiparticles screening the time-dependent deformations of the CDW. For this reason, the CDW conductivity is proportional to the conductivity of quasiparticles and decreases with decreasing temperature \cite{sigmaCDW1,sigmaCDW2,sigmaCDW3}.

4) $E\gg E_T$. In this area, so called “Fr\"ohlich superconductivity” arises  \cite{FrSC}. Current-voltage characteristics  ({\it IV } curves) are almost vertical in this regime. In our opinion, this mode arises when the CDW sliding frequency (i.e. narrow band noise frequency, that is equal to the inverse CDW travel time by one period) exceeds the reverse quasiparticle Maxwell relaxation time. Magnetic fields of up to 16 T do not affect this phenomenon in the topologically trivial blue bronze K$_{0.3}$MoO$_3$ \cite{FrSC}.

The results presented in Ref.~\cite{gooth} 
indicate that under certain conditions the CDW kinetics (magnetic-field dependent) is independent of the kinetics of quasiparticles (magnetic-field independent), i.e. a scaling violation occurs.

Here we present the results of study of MR in the q-1D conductor (TaSe$_4$)$_2$I, performed in wider ranges of temperatures and electric fields. Our aim is to explore conditions required for observation of such a violation. We found that the longitudinal MR in the region of nonlinear conductivity in the magnetic fields up to 7 T is positive, does not exceed a fraction of a percent and is proportional to the longitudinal MR in the linear conductivity region. However, under conditions of spatially inhomogeneous motion (in samples profiled by focused ion beams), negative MR is absent in the region of linear conductivity but appears in the region of nonlinear conductivity, that indicates scaling violation. The possible contribution of dislocations and solitons in observed MR is discussed. 

\begin{figure}
\includegraphics[width=7cm]{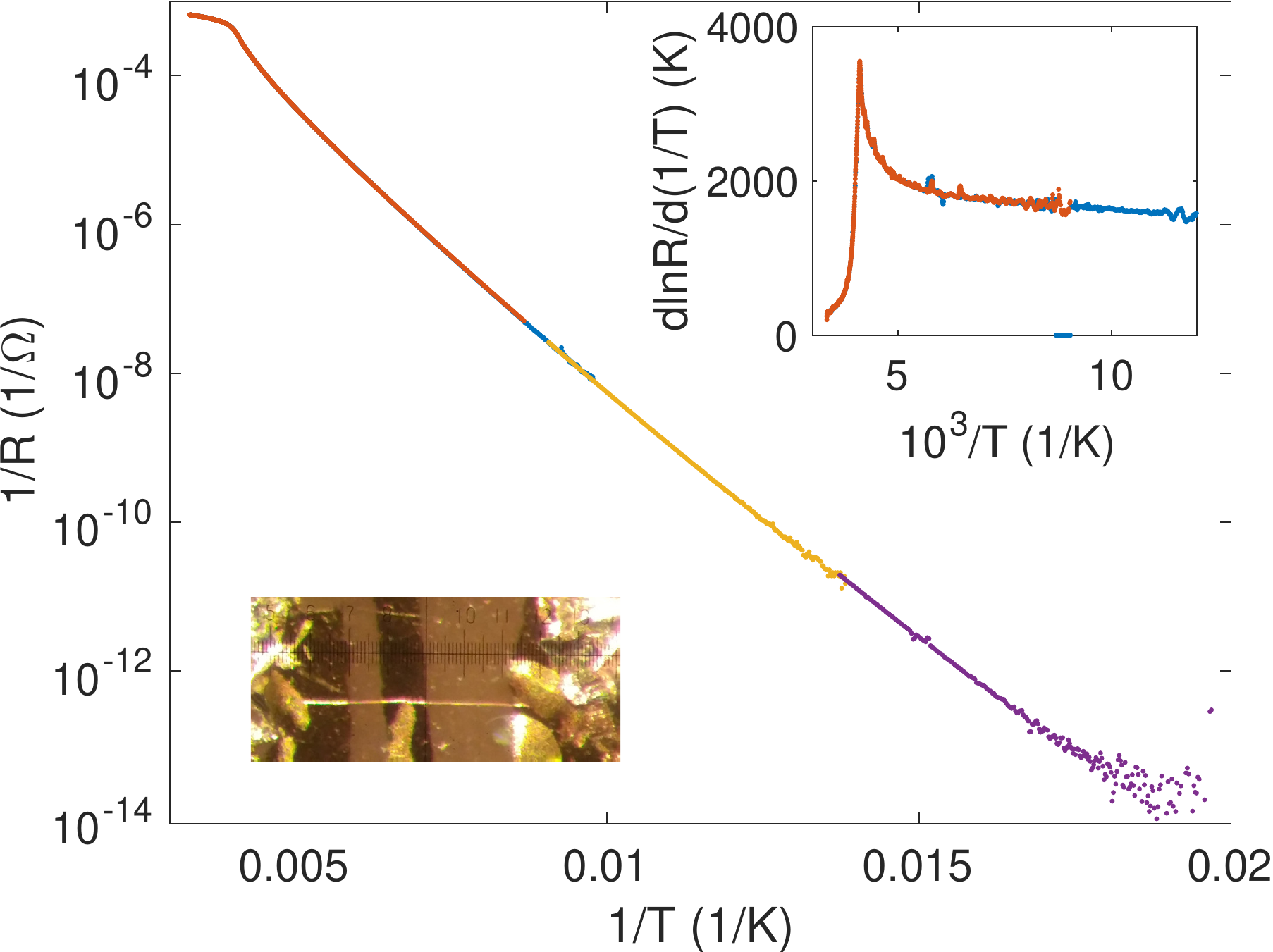}
\caption{Temperature dependent linear conduction in (TaSe$_4$)$_2$I (long segment). Left inset: sample image. Right inset: logariphmic derivative of the conduction curve. }
\label{rtplot}
\end{figure}

We study samples (TaSe$_4$)$_2$I of the same origin as those studied in Refs.~\cite{termoEDS,s1,s2}. The data presented in this work were obtained on two segments of a relatively thin sample with transverse dimensions $w\times t=4.0 \times 1.9 $ $\mu$m$^2$ and distances between the edges of the contacts $L_s=130$ $\mu$m  and $L_l= 330$  $\mu$m (see inset in Fig. \ref{rtplot}). The use of short and thin samples having 2 orders of magnitude smaller cross-sectional area than those in \cite{gooth} allows measurements of nonlinear conductivity at much higher values of the electric field due to better heat exchange intrinsic to thin samples. At the same time, the transverse dimensions of the studied samples were not so small as to lead to the appearance of finite-size effects, which can significantly affect collective transport \cite{zzreview}. 

We use vacuum deposited gold contacts providing a low contact resistance of the order of 1-10 $\Omega$ \cite{forro}.  It turns out that {\it IV} curves of the long and short segments, measured by the two-contact method, are similar to each other with a scaling factor of $R_s/R_l=0.42$ (see also below). A slight difference between the ratios $L_s/L_l=0.39$ and $R_s/R_l=0.42$ is due to the presence of a geometric contribution to the contact resistance (spreading resistance), which leads to an increase in the effective length of the sample by $\sim t\sqrt{A}$, which in our case amounted to 15 $\mu$m. We also performed $R(T)$ measurements in the four-probe configuration with cold-soldered indium contacts. 
The comparison of the results obtained in two- and four-probe configurations also indicates a negligible contribution of the gold contact resistance to the measured effects. The use of low-resistance contacts makes it possible to carry out all measurements using the two-probe method, which, in turn, allows us to advance to the region of significantly lower temperatures than in previous studies. All the measurements reported below were carried out in the voltage controlled mode.

The room-temperature resistivity of the samples studied, estimated from the above data, is $2.8\times 10^{-3}$ $\Omega\cdot$cm, practically coincides with the value $2.9\times 10^{-3}$  $\Omega\cdot$cm from \cite{makigruner}, and is about twice larger than $1.5\pm 0.2\times 10^{-3}$ $\Omega\cdot$cm reported in \cite{wangmonceau,gooth}. 

Temperature-dependent conduction of studied samples is shown in Fig. \ref{rtplot}. The dependence has the usual form for (TaSe$_4$)$_2$I and is characterized by semimetal-like temperature-dependent conductivity around room temperature, and the Peierls transition leading to development of a low-temperature gapped state at $T<T_P$. The inset shows the derivative $d\ln (R)/d(1/T)$, which is the instant activation energy. The maximum of the derivative corresponds to the Peierls transition temperature $T_P=245$ K. The activation energy is 1650 K at 100 K and gradually decreases with decreasing temperature. Such a decrease is characteristic of many q-1D conductors with CDW \cite{CDWreview}. In general, the temperature dependences of the conductivity of the studied samples correspond to those obtained previously \cite{termoEDS,wangmonceau,makigruner}.

\begin{figure}
\includegraphics[width=8cm]{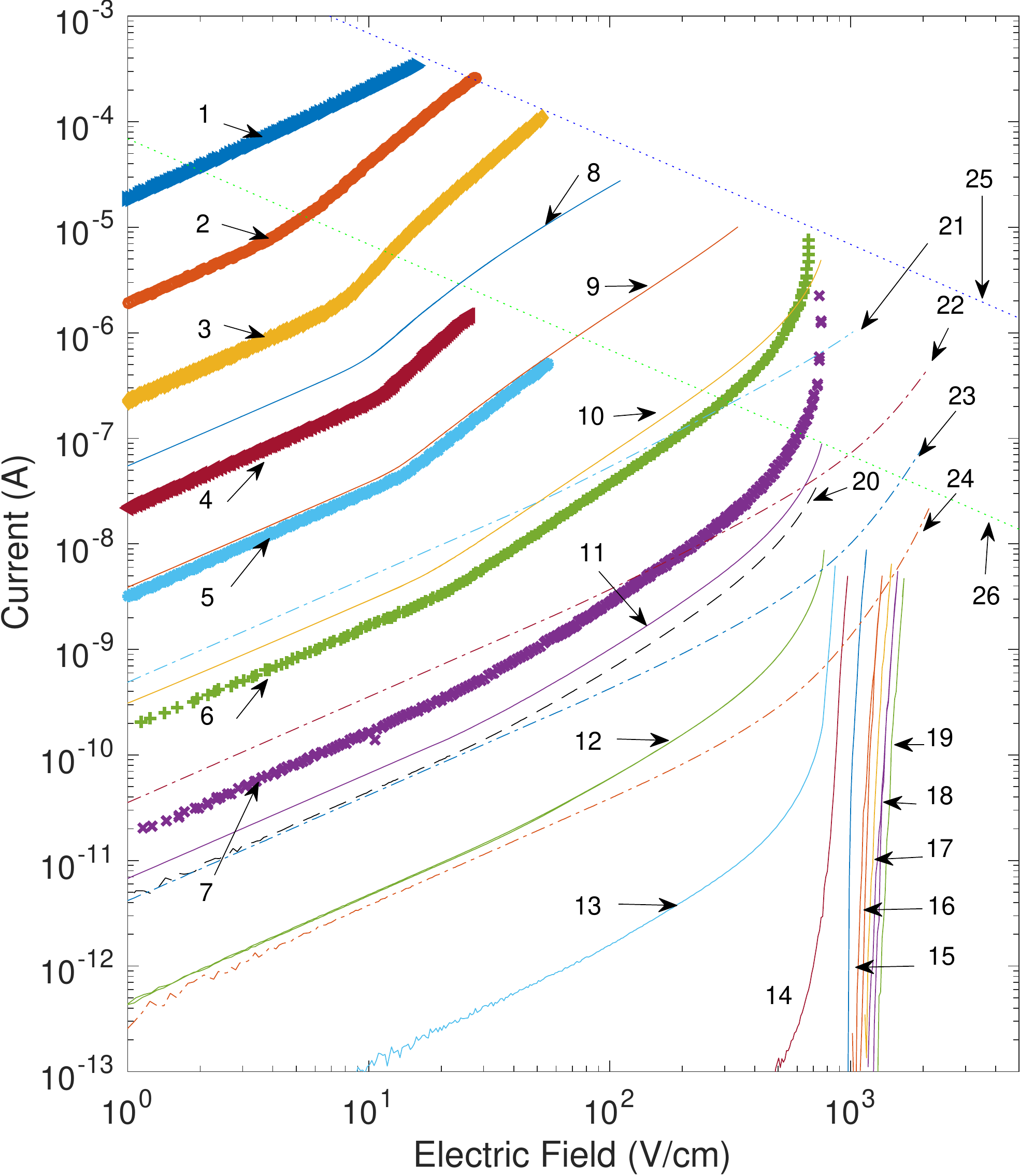}
\caption{Temperature set of {\it IV} curves in (TaSe$_4$)$_2$I. (1-7) - $IV$ curves of the long segment measured at temperatures of (1) - 250 K, (2) - 207 K, (3) - 167 K, (4) - 140 K, (5) -119 K, (6) - 96 K and (7) - 87 K. (8 -19) - $IV$ curves of a short segment measured at temperatures of (8) - 150 K, (9) - 120 K, (10) - 100 K, (11) - 80 K, (12) - 70 K, (13) - 60 K, (14) - 50 K, (15) -40 K , (16) - 30 K, (17) - 25 K, (18) -20 K, (19) - 15 K. (20) - $IV$ curve of an E-type sample at $T= 80$ K. (21-24) - $IV$ curves of a W-type sample measured at temperatures (21) - 120 K, (22) - 100 K, (23) - 80 K, (24) - 70 K. (25-26) are the boundaries corresponding to the power of (25) - 100 $\mu$W and (26) - 1 $\mu$W in the short segment. }
\label{replot}
\end{figure}

Fig.~\ref{replot} shows a temperature set of the {\it IV} curves of the long (curves 1-7) and short (curves 8 -13) segments of the sample. 
Matching the {\it IV} curves of short and long segments proves the absence of a significant contribution of the contacts to the linear and nonlinear conductivity. The {\it IV} curves obtained in magnetic fields  3.77 and 7.14 T  at temperatures 60, 80, 100 and 120 K are practically undistinguishable from zero-filed ones (see below) and are not shown in this figure. 
 
\begin{figure}
\includegraphics[width=7cm]{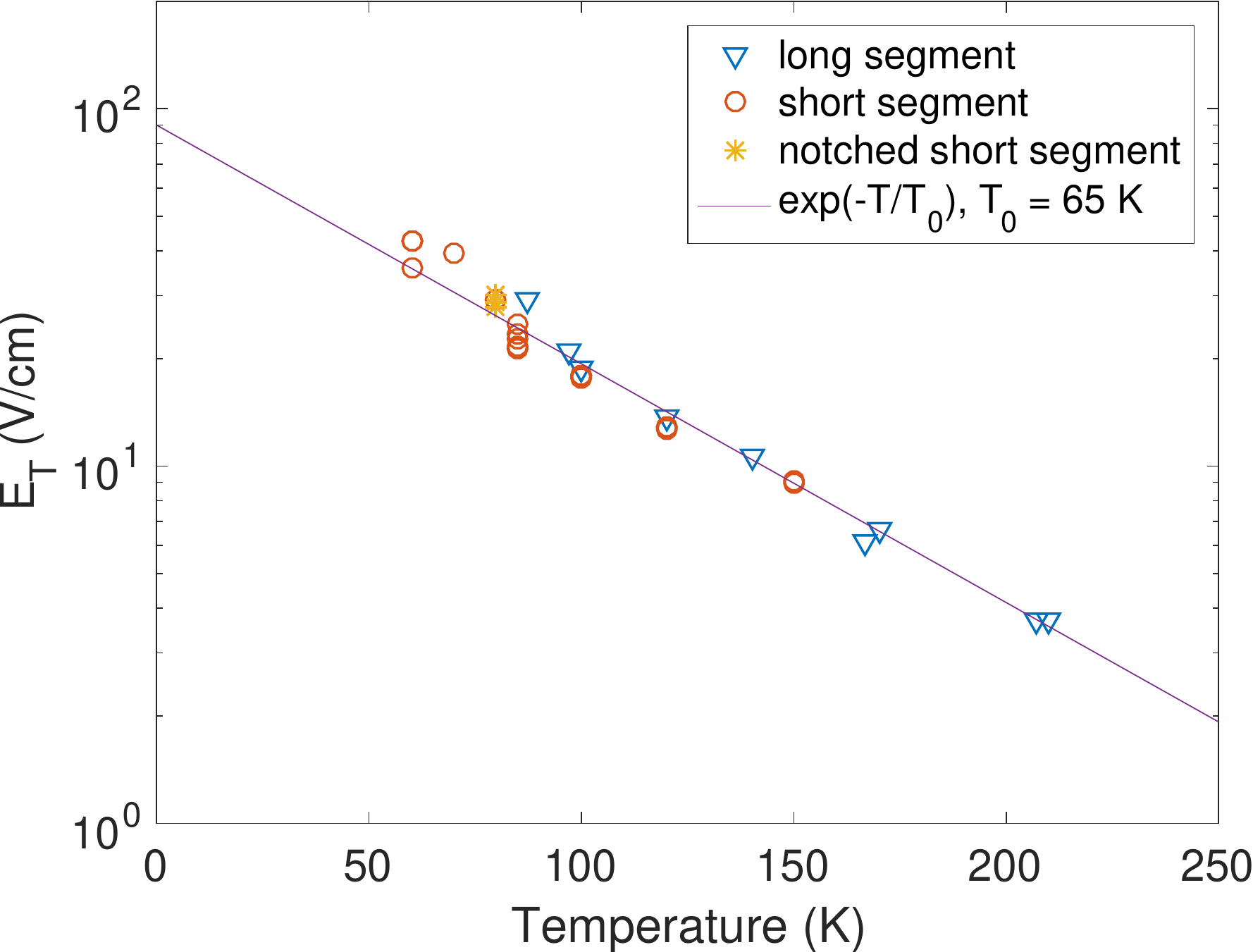}
\caption{Temperature dependence of the threshold field for two segments of (TaSe$_4$)$_2$I.  }
\label{etplot}
\end{figure}

Fig.~\ref{etplot} shows the temperature dependence of the threshold field, which in this work was determined by the criterion of a 3 \% conductivity increase with respect to the linear one. This dependence corresponds to the law $E_T\propto \exp{(-T/T_0)}$ typical for Peierls conductors including  (TaSe$_4$)$_2$I \cite{EtT} and associated with temperature fluctuations of the order parameter \cite{Maki}. 

At $E< 10$ V/cm, our results of the study of linear and nonlinear conductivity coincide with ones obtained previously  \cite{termoEDS,wangmonceau,makigruner} and correspond to the usual behavior of q-1D conductors with CDWs. In addition, a region of Fr\"ohlich superconductivity \cite{FrSC} is found to develop at $E\sim 10^3$ V/cm which was not observed ealier in this material.  The results fit into the general picture of phenomena observed in the Peierls conductors that was described in the introduction. 

\begin{figure}
\includegraphics[width=8cm]{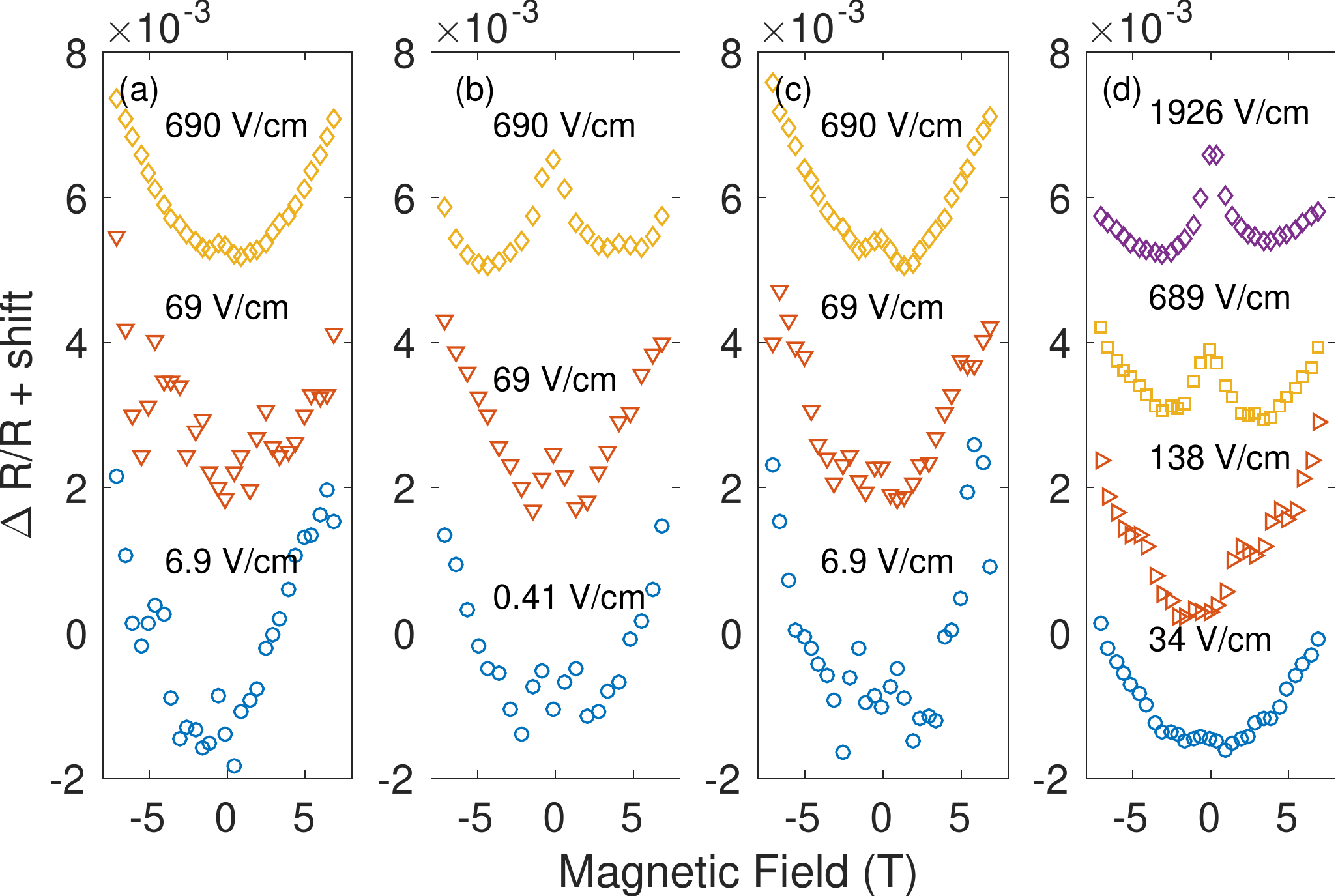}
\caption{Magnetoresistance in (TaSe$_4$)$_2$I in different conduction modes, sample and measurement geometries at $T=80$~K: a) Initial sample, $B\parallel I$; b) Initial sample, $B\perp I$, $E_T= 30$ V/cm. c) One-side notched sample (E-type sample), $B\parallel I$, $E_T= 30$ V/cm. d)Two-side notched sample (W-type sample), $B\parallel I$, $E_T=140$~V/cm. 
}
\label{magres}
\end{figure}

Fig.~\ref{magres}(a) shows the normalized longitudinal MR $\Delta R /R\equiv \left[R (B)-R(0)\right]/R(B)$, where $B$ is magnetic field, measured in different regions of the {\it IV} curves. The MR is positive and quadratic, and its magnitude does not exceed one percent. As expected, the relative changes in linear and nonlinear conductivity are proportional to each other in all parts of the {\it IV} curves within the measurement accuracy. We see that CDW transport in our samples is not accompanied by the negative longitudinal MR.

The transversal MR is shown in Fig.~\ref{magres}(b). The region of negative MR is clearly seen in the nonlinear conduction regime but is absent in the linear one (see also below). This type of behavior represents another example of violation of  scaling between the linear conduction and CDW damping, as it is observed under another orientation of the magnetic and electric fields than in Ref.~\cite{gooth}.

We may assume that the absence of the scaling in Ref.~\cite{gooth} may occur due to a deviation from conditions under which the kinetics of CDW is usually studied. 
Indeed, since in \cite{gooth}, the power released in the sample was large (up to 60 mW in the region of the effect, instead of 1 $\mu$W in the present measurements), we can expect the appearance of an inhomogeneous temperature distribution over the sample cross-section.  Since $E_T$ exponentially depends on temperature (see Fig. \ref{etplot}), the velocities of CDW sliding in the center of the sample cross-section and at its edges are different. Such a difference should be accompanied by appearance and gliding of CDW dislocations. 
As the CDW order parameter is suppressed in the dislocation core, so the properties of the Weyl semimetal, which are characterized by negative MR, could partially restore. 

\begin{figure}

\includegraphics[width=4.4cm]{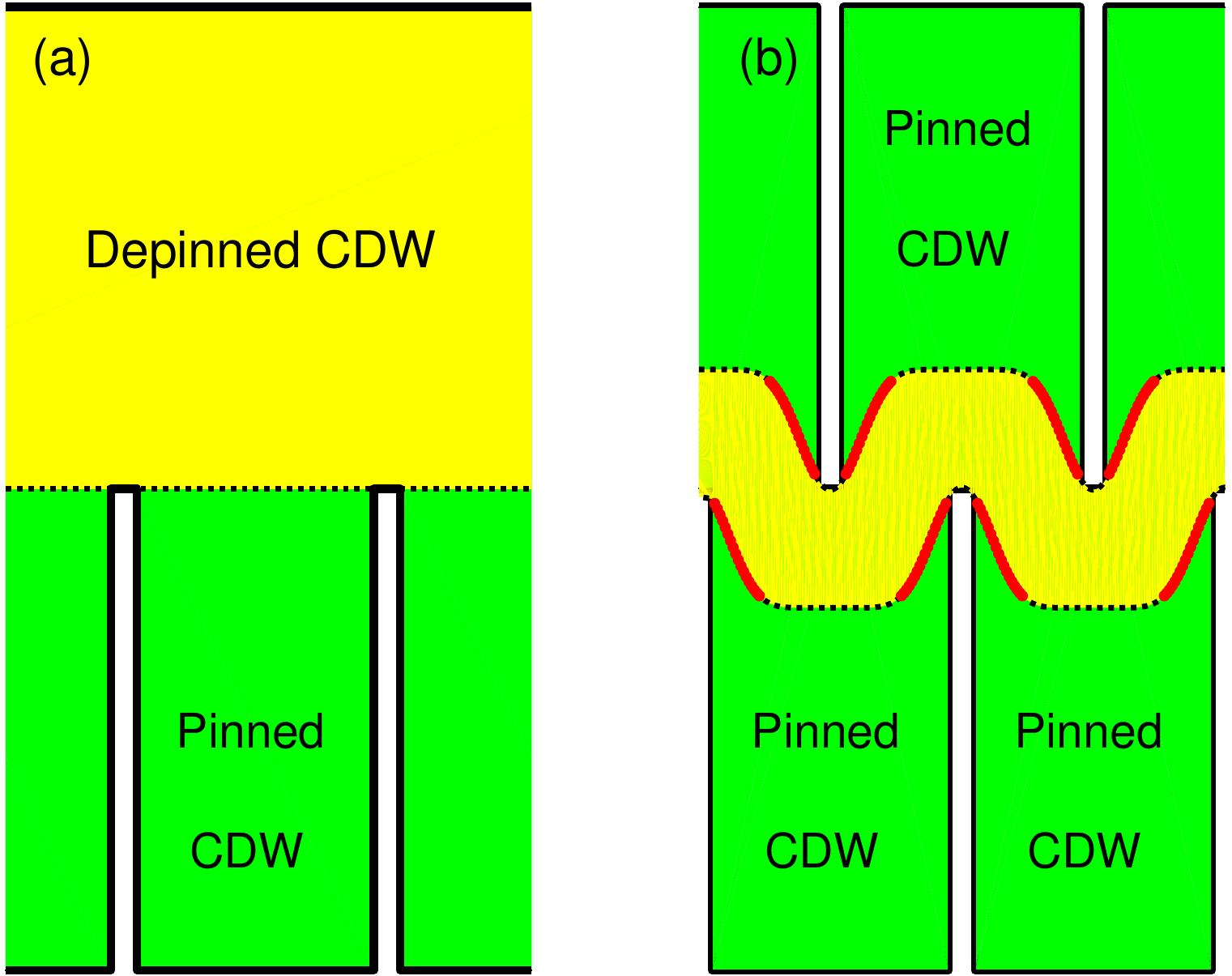}
\ \ \
\includegraphics[width=3.7cm]{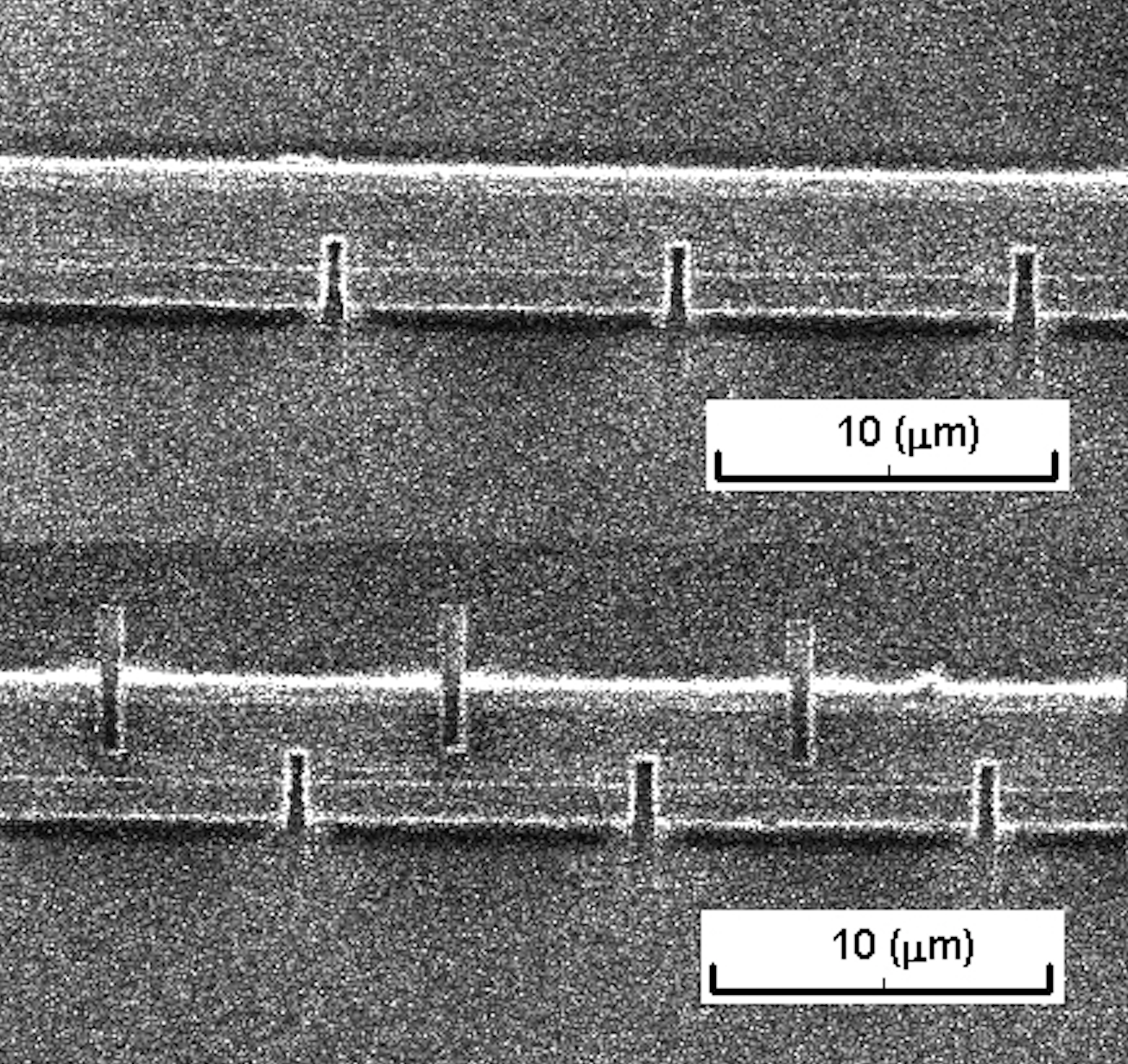}
\caption{Schemes of fragments of (a) single-side notched  sample (E-type) and (b) double-side notched samples  (W-type) in the isotropic representation. Regions with pinned  and sliding CDW are marked by green and yellow colors respectively. Regions of phase slip are marked in red. c)  Images of the FIB etched sample.}
\label{sample1}
\end{figure}

To study a possible contribution of dislocations into CDW kinetics, we modified the sample geometry. Namely, by using FIB, a set of cuts was made from the edge to the sample center (Fig. \ref{sample1}(a)). The distance between the cuts $l=10$ $\mu$m was enough  to provide pinning in the notched segments of the sample, $w\sqrt{A}/2 > l$. In such a sample (we call it E-type sample), CDW sliding is accompanied by motion of dislocations along the boundary between pinned and depinned CDW. 

An {\it IV} curve of the E-type sample is shown in Fig. \ref{replot} (curve 20). The conductance of the E-type sample is smaller by 1.48 times in the linear mode and 1.57 in the nonlinear one with respect to the initial values (curve 11). A stronger decrease in the nonlinear mode indicates that, indeed, the nonlinear current is now non-uniformly distributed across the sample, and thus CDW sliding is accompanied by motion of CDW dislocations. However, the longitudinal MR in the E-type sample remains roughly the same as before cutting (Fig. \ref{magres}(c)).  

\begin{figure}
\includegraphics[width=7.5cm]{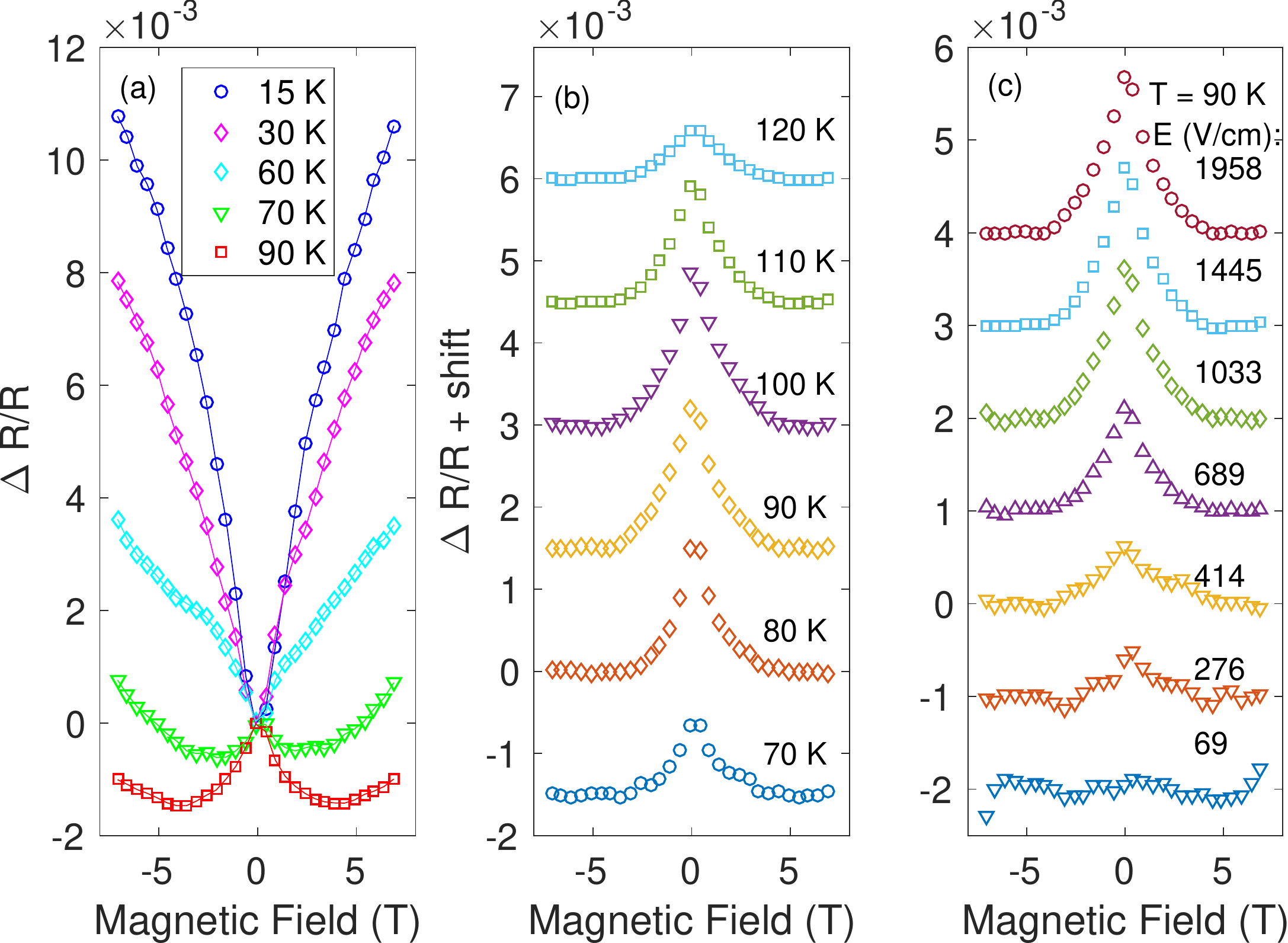}
\caption{a) Temperature evolution of the longitudinal MR.  Evolution of the negative MR contribution with temperature (b) and the electric field (c).  W-type sample.}
\label{PMR}
\end{figure}

To increase the contribution of dislocations, additional cuts were made from the opposite side (Figure \ref{sample1}(b)). It was assumed that the sliding of the CDW in such a sample with notches on both sides (a W-type sample) will be strongly inhomogeneous and be accompanied by phase slips at the boundary between the pinned and sliding CDWs (red lines in the figure \ref{sample1}(b)). The $IV$ curves of a W-type sample are shown in the Figure \ref{replot} by dashed-dotted lines (curves 21-24). It turns out that such incisions from both sides, indeed, lead to the appearance of a negative MR in nonlinear conductivity region (Figure \ref{magres}(d)). As can be seen from this figure, in the W-type sample, negative MR develops against the background of positive. Both contributions are temperature dependent.  At $T \lesssim 30$ K, only the nonlinear conduction region is available for study. With decreasing temperature, parabolicity disappears (Figure \ref{PMR}(a)).  The observed MR corresponds to weak antilocalization of gapped electrons in the case of strong spin-orbit interaction \cite{locantiloc}. At temperatures $T <70$ K, no negative MR regions observed. With decreasing temperature, the positive MR increases by almost an order of magnitude and reaches 1.6\% at $T = 6$~K.
  
 $\Delta R(B)/R(0)$ dependences obtained after subtraction of the positive parabolic MR at $T> 70$ K are shown in Fig. \ref{PMR}(b). Separation of the negative and positive magnetoresistance contributions al lower temperatures is problematic because of experimental unavailability of the linear conduction contribution, dramatic change of the shape of $\Delta R(B)/R(0)$ dependence and increase of its value (Fig. \ref{PMR}). 
 Such a negative MR contribution exists at $E>E_T$ only and increases with the CDW current (Fig. \ref{PMR}(c)). No negative MR region is observed at temperatures $T\lesssim 60$~K. 
 
In the W-type sample, the transverse MR remains almost the same as in the initial sample (Fig. \ref{magres}(b)). In the linear conduction regime, no negative MR region is observed in either orientation.
 
 $\Delta R(B)/R(0)$ dependences shown in Fig. \ref{PMR}(a) are typical for weak antilocalization ($T< 60$~K) and both weak localization and antilocalization ($T\geq 70$~K) respectively \cite{loc,locantiloc,locreview}. Such a complex weak localization-antilocalization behavior was earlier observed in strained InGaAs/InP quantum well structures \cite{stud}. A feature of the obtained results is the appearance of the negative MR in the nonlinear conduction modes only. 
 
 As it was noted above, sliding of the CDW is accompanied by phase slips of the CDW on the boundary between its pinned and sliding regions \cite{loops,disl,ps,braz}. There are two main scenarios of phase slip in the Peierls conductors. The first one consider phase slip due to nucleation of dislocation loops (that are channels for quasiparticle motion) of a critical diameter \cite{loops,disl,ps}.  Another scenario considers phase slip as a sequence of elementary process: injection of electrons from the contacts, their transformation into solitons, and aggregation of solitons into complexes that are equivalent to dislocation loops \cite{braz}. It seems that dislocation loops by itself have no relation to the transverse negative MR in the initial sample (Fig. \ref{magres}(b)), but solitons may have. Indeed, if some of them remain unaggregated, they provide an additional conduction. As jumps of solitons across the chains are less probable than their motion along, the closed-loop trajectories oriented along the chains are more probable than the transverse one. Respectively, the weak localization in the initial sample is expected to be  more pronounced in the transverse MR than in the longitudinal one, in agreement with the experimental results (compare Figs. \ref{magres}(a) and \ref{magres}(b)). Alternatively, the difference between MR in the linear and nonlinear conduction regimes may result from a change in MR of quasiparticles.  Further study is required to clarify the origin of this difference.  Appearance of the longitudinal negative MR in the W-type sample corresponds to mixing of the longitudinal and transverse contributions due to complex current trajectory.

The negative MR observed in the W-type sample is much smaller than in ~\cite{gooth}. In the case of a chiral anomaly, the change in conductivity in a longitudinal magnetic field depends on the level of the chemical potential $\mu$ as $\delta \sigma \propto \mu^{-2}$, where $\mu$ is measured from the apex of the Weyl cone \cite{Son,Burkov}. The crystals studied in this work have 7\% lower $T_P$, and 1.5-2 times higher values of $E_T$ and the fluctuation parameter $T_0$  than their respective values in \cite{gooth}.
In (TaSe$_4$)$_ 2$I, these differences may be due to a violation of the stoichiometry for  iodine \cite{Gressier}, which leads to a change in $\mu$  and, therefore, temperature dependence of the concentration of current carriers $n(T)$ in the semimetal phase. The difference in these dependences is indicated by different slopes of the $R(T)$ curves for $T>T_P$. So, at a temperature $T = 1.14 T_P$ for the samples studied in this work, the value $d\ln R/d\ln T |_{T = 280 \rm {~ K}}= - 1.5$, and for the samples from the works \cite{makigruner, gooth} $d\ln R/d\ln T|_{T = 300 \rm {~ K}} \approx -3$. The difference in these values makes it possible to roughly estimate the difference in the position $E_F$ of these samples, which is 40-70 meV, depending on the details of the band structure \footnote{Using the difference of the slopes allows us to get rid of the contribution of the temperature dependence of mobility. To estimate $n(T)$, data on the position of 48 Dirac cones (table I of Ref. \protect\cite{shi}), as well as data on the band structure of Ref. \protect\cite{zhang,li} were used. The calculation are performed according to the formula $n(T) \propto\sum_i \left[{\int_{E_i}^{\infty}D_i (E) / (1+e^\frac{E-E_F}{kT})dE} + {\int_{-\infty}^{E_i}D_i (E)/(1 + e^{{\frac{E_F-E}{kT}}}}dE\right]$, where $D_i \propto | E- E_i |^{d_i}$ is the density of states for the $i$-th cone of dimension $d_i$ with an apex sitting at an energy of $E_i$. The results turns out to be weakly dependent on the initial data, assuming that $E_F$ in \protect{\cite{gooth}} is located near the apexes of the cones.}.  The disorder caused by impurities (in this case, iodine vacancies) also reduces the effect \cite{disorder1, disorder2}. In addition, impurities can even change the functional dependence from parabolic to the linear one \cite{linmag}.

In conclusion, 
we demonstrate here that spatially nonuniform motion of the CDW in the Weyl semimetal 
(TaSe$_4$)$_2$I reveals rich physics of both CDW and topological materials.  Namely, we show that  the CDW kinetics may manifest the negative magnetoresistance of the weak localization type.  This negative magnetoresistance is developed on the background of the positive magnetoresistance corresponding to antilocalization inherent to topological materials. The results of this work indicate the key role of experimental conditions and sample parameters for observing the chiral anomaly in the Peierls state.

{\bf Acknowlegements.} We are grateful to Helmuth Berger for providing the samples and N.I. Fedotov for useful comments. I.A. Cohn and S.V. Zaitsev-Zotov acknowledge financial support from the RSF (grant \#16-12-10335), S.G. Zybtsev and A.P. Orlov performed sample preparation and their profiling by FIB in the framework of the Russian State task.

\end{document}